# Video Screens for Hearing Research: Transmittance and Reflectance of Professional and Other Fabrics

Jan Heeren[1,2,4], Giso Grimm[2,3], Stephan Ewert[2,3], Volker Hohmann[2,3]

[1] *Hörzentrum Oldenburg gGmbH, Oldenburg*
[2] *Cluster of Excellence Hearing4all, Oldenburg*
[3] *Carl-von-Ossietzky-Universität Oldenburg*
[4] *Email: heeren@hz-ol.de*

## Abstract

Virtual reality labs for hearing research are commonly designed to achieve maximal acoustical accuracy of virtual environments. For a high immersion, 3D video systems are applied, that ideally do not influence the acoustical conditions. In labs with projection systems, the video screens have a potentially strong influence depending on their size, their acoustical transmittance and their acoustical reflectance. In this study, the acoustical transmittance and reflectance of six professional acoustic screen fabrics and 13 general purpose fabrics were measured considering two tension conditions. Additionally, the influence of a black backing was tested, which is needed to reduce the optical transparency of fabrics. The measured transmission losses range from -5 dB to -0.1 dB and the reflected sound pressure levels from -32 dB to -4 dB. The best acoustical properties were measured for a chiffon fabric.

## Introduction

In hearing research, virtual reality (VR) labs aim for closing the gap between the realism of field studies and the control and reproducibility of lab studies. The main purpose is to assess the listening performance or behavior of human subjects and hearing devices in realistic sound environments. Room acoustics has a critical influence on psychoacoustical measures, e.g. localization [1] and speech perception [2]. To achieve plausible outcomes in virtual environments, it is crucial that the measurement room and the setup have minimal influence on the listening conditions. In VR labs with large projection systems, the video screen is commonly the major surface in the setup. A high reflectance of the screen fabric would have a major impact on the listening conditions, particularly in setups with curved screens or opposing parallel screens as in "cave"-like arrangements [3]. In the literature reflected sound pressure levels (SPL) between -15 and 0 dB are reported for professional video screens [4] typically used in, e.g., cinemas.

Additionally, such labs need loudspeaker setups that fulfil certain technical and psychoacoustical requirements [5]. As the loudspeakers have to be placed behind the video screen, this sets a second requirement to the screen fabric: maximal acoustical transmittance. The energy of the transmitted sound is reduced due to the reflection and the absorption effects at the screen [6]. While the direct transmitted sound from a loudspeaker can in principle be equalized for a certain direction, multiple transmission paths can occur.

The acoustical characteristics of fabrics strongly depend on the material, the density, and the structure [4, 7, 8]. A basic impedance model for fibrous materials was published by Delany and Bazley [9] that describes the frequency dependency of transmitted sound as a function of the flow-resistance. For thin layers < 2 cm the frequency response has a low-pass characteristics with a cut-off frequency > 1 kHz [9]. Actual acoustic screen fabrics show rather higher cut-off frequencies [4, 10]. Considering only fabrics that are usable as a projection screen, transmission loss values in the literature range from -5 dB to -2 dB [4, 7].

In this study, 19 fabrics have been evaluated regarding transmittance and reflectance. Some of them are professional acoustic video screen fabrics and some are decoration fabrics or cardigans. None of the professional video screens came with reference data for reflectance, but the retailers provided transmittance specifications. The tested fabrics vary regarding optical parameters. Although this study does not include an optical evaluation, one very common optical optimization tool is considered in the acoustical evaluation: a black backing. This is used to reduce the optical transparency of the screen and particularly to avoid the projection being visible on surfaces behind the screen [12]. To consider potential effects of differences in elasticity across the different fabrics, two tension conditions were measured.

## Method

### Tested fabrics

Four providers of professional video screens, one big fabric distributor and two local fabric stores were requested to provide 1 m x 1 m samples of all fabrics that meet the following requirements:

1. white colored (uni, any shade)
2. smooth surface
3. low weight

In total 19 different fabrics were tested as listed in Table 1.

**Table 1:** Tested fabrics

| Style | Material | Transmittance specification |
|---|---|---|
| Chiffon | polyester | |
| Lycra | nylon/spandex | |

| Jersey | cotton/spandex | |
| Power stretch | polyester | |
| Mesh | polyester | |
| Stage mesh | polyethylen | |
| Taft, cream | polyester | |
| Taft, snow | polyester | |
| Taft, cream | acetat/polyamid | |
| Bunting, cream | cotton | |
| Bunting, snow | cotton | |
| Canvas, cream | cotton/polyester | |
| Canvas, snow | cotton/polyester | |
| Professional 1 | unknown | -3 dB |
| Professional 2 | unknown | -2.2 dB (2-20 kHz) |
| Professional 3 | unknown | -3 dB (0.3-20 kHz) |
| Professional 4 | unknown | -1.5 dB (2-20 kHz) |
| Professional 5 | unknown | -1.8 dB (5-30 kHz) |
| Professional 6 | polyester/vinyl | -4 dB maximal difference (7 kHz) |
| Black backing | unknown | -0.25 dB (10 kHz) |

### Setup

Measurements were conducted in a sound treated room with dimensions 4 m x 4 m x 2.6 m (length, width, height). Figure 1 shows a sketch of the measurement setup. Fabric samples of 1 m x 1 m were clamped between pairs of wooden bars at the top and the bottom edges. The upper pair of bars was hung up below the ceiling using four cord lines of 0.5 m length. The lower pair of bars was weighted with four variable iron weights that were also tied using cord lines (same length). Weights of either 1.25 kg or 0.75 kg (3 kg or 5 kg total weight) were used. On one side of the screen a loudspeaker (Genelec 8030B) was placed (0.6 m distance, 0.1 m left of the screen center). A microphone (Neumann KM 183) for reflection measurements was placed beside the loudspeaker (same distance, 0.1 m right of the screen center). The distance of 0.12 m between the loudspeaker and the microphone is needed to avoid distortions of the polar pattern of the microphone. For transmission loss measurements, another KM 183 microphone was set up on the other side of the screen (same distance, center position). All other surfaces and objects in the room had sufficient distances not to produce disturbing reflections. In the test condition with black backing, the backing fabric was attached on the loudspeaker side without a gap to the screen fabric. Due to the tension, some fabrics showed minor ripples in shape. These caused some variance in the gap between 0 and 3 cm. When setting up a screen, it was made an effort in avoiding ripples. If this was not perfectly feasible, it was decided to proceed with ripples and consider this fabric characteristic this way.

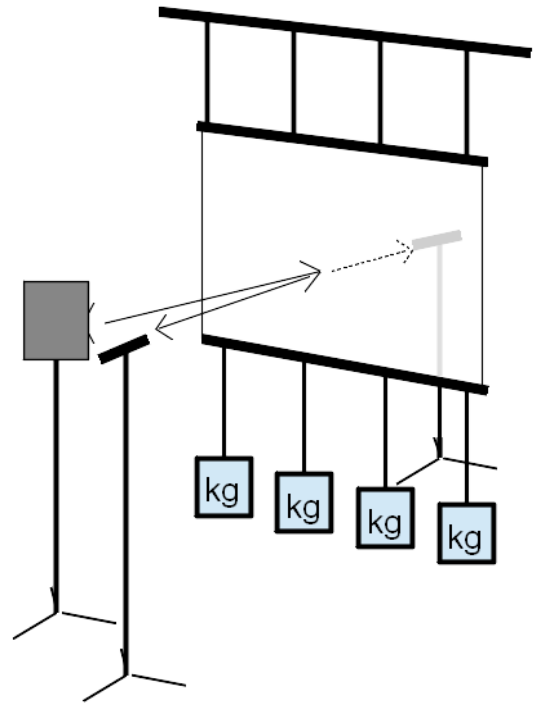

**Figure 1:** Sketch of the test setup.

### Measurement of transmission loss and reflected SPL

Impulse responses (IRs) were measured based on sine sweeps according to [11]. Logarithmic sweeps with a frequency range of 50-18000 Hz and a duration of 2^14 samples were applied. The sample frequency was 44100 Hz.

As a reference, an IR was recorded without a fabric sample. For this purpose, cord lines were used to hold the lower bars that were clamped in at the left and right edges, so that there is no obstacle in the plane of the screen for this condition. The reference IR recorded with the transmission microphone is the reference for both transmission and reflection IRs. For the analysis, the time frame from 2.9-4.1 ms after playback was windowed applying a von-Hann-window. This time frame is limited by the reflection from the wood bars. Correspondingly, the measurement covers a frequency range from 950-18000 Hz, which is the most relevant range for the expected effects [4, 10]. Transmission loss and reflected SPL are given by the difference in power between the test and the reference condition.

### Results

Figure 2 shows an overview of the measured reflectance and transmittance of all 19 fabrics. The professional acoustic screen fabrics are presented in red and the other fabrics in grey. Reflected SPL values range from -32 dB to -4 dB. Some show a relatively flat frequency response (+-2 dB), others show variances or high pass characteristics with slopes of up to 3 dB per octave (12 dB over the presented range). The professional fabrics cover a range from -22 dB to -4 dB.

Transmission losses range from -6 dB to 0 dB. Generally, the transmission spectra show a low-pass characteristic. Some fabrics show ripples in the spectrum. In a pre-test, these ripples were different. The ripples occurred for fabrics that become wavy under tension.

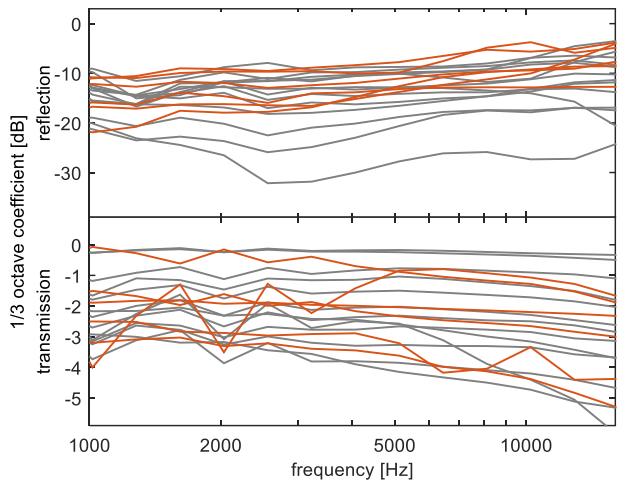

**Figure 2:** Overview; reflectance and transmittance of six professional acoustic screen fabrics (red) and 13 other fabrics (grey).

In figure 3 data for a selection of the three best performing fabrics (chiffon, mesh, and lycra) and the best professional fabric (professional 4) is shown. Data measured with 5 kg tension weights are plotted with solid lines, data for the 3 kg condition are shown by dotted lines. The measured values for the two weight conditions are almost equal with differences of 0.5 dB in reflected SPLs and 0.1 dB in transmission losses.

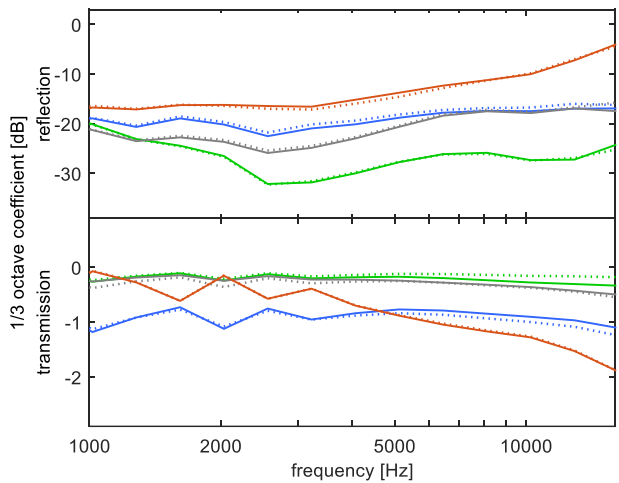

**Figure 3:** The three best performing non-professional and the best professional screen fabric are shown; reflectance and transmittance of chiffon (green), mesh (grey), lycra (blue) and professional 4 (red); solid line values were measured with 5 kg tension weight, dotted lines with 3 kg.

In figure 4 the solid lines show the same data as in figure 3. Different from figure 3, the dotted lines show data of the black backing condition. For fabric "professional 4" the black backing condition shows hardly any difference in reflected SPLs from without black backing. For the other three fabrics the black backing shows a decrease in the range of 3-6 kHz of up to 3 dB and an increase around 9-16 kHz of up to 8 dB. Transmittance values are influenced very evenly over all frequencies. They are decreased by 0.1-0.3 dB for the black backing condition.

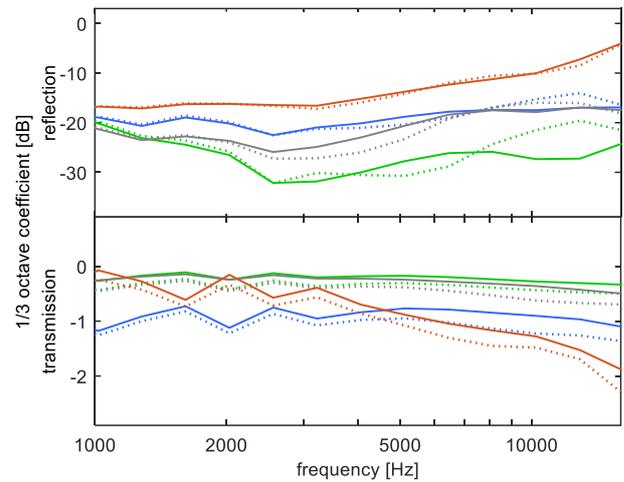

**Figure 4:** The influence of a black backing; reflectance and transmittance of chiffon (green), mesh (grey), lycra (blue) and professional 4 (red); solid line values were measured without, dotted lines with a black backing.

## Discussion

Nine of the tested fabrics show much lower reflected SPLs than the fabrics measured by Long et al. [4]. Values comparable with the Long study [4] were only measured for the ten worst performing fabrics, including five of six professional screens. The best performing fabrics in this study show reflectance values of down to -32 dB.

The measured transmission losses are comparable to the specifications provided by the retailers and reproduce literature values for woven fabrics [4, 13]. "Professional 4" even performed better than the retailer indicated, whereas the results of "Professional 6" are dominated by ripples caused by waviness. Pre-tests have proven that ripples in the frequency responses between 1-3 kHz are evidence for waviness, that occurs in some fabrics if they are mounted under tension.

With transmission losses between -2 dB and 0 dB the most interesting fabrics (selection shown in figure 3 and 4) have a very low influence on transmitted sound. Chiffon and mesh do not even exceed -0.5 dB. Furthermore, the frequency responses are flatter than the data published by Long et al. [4]. According to the Delany and Bazley model [9], this indicates that the lowpass-filter characteristics simply show cut-off frequencies higher than 20 kHz. In the case of chiffon the extremely low weight of the fabric (77 g/m²) corresponding with a low density is remarkable and may make the difference.

Subjectively judged, these four fabrics also cover the complete range of visual quality that may be expected from a potential screen fabric. Chiffon has the highest optical transparency and thus may not be suitable for applications that require a high brightness. Its structure is fine and plain, though, which makes it suitable for high resolution projections. The mesh fabric and professional 4 can be stated as medium bright, while professional 4 may be a little better for higher resolutions. The lycra sample is shinier and can be used to reach high brightness values. It also has a fine structure and can be used for high resolutions.

## Conclusion

Reflectance and transmittance of six professional video screen fabrics and 13 general purpose fabrics were tested. Six fabrics showed an excellent transmittance and three fabrics showed very good reflectance values. The three acoustically best performing fabrics were not professional video screen fabrics. The acoustical properties of the chiffon fabric were outstanding compared with all other fabrics in the test. With respect to visual factors, also the second and third best fabrics as well as the best performing professional fabric may be suitable for hearing lab applications. A comparison of two tension conditions hardly showed any effect. Adding a black backing does have a minor influence on the test measures while the choice of the fabric is the dominant factor.

## Acknowledgments

This work was supported by the Deutsche Forschungs-Gemeinschaft (DFG, German Research Foundation) – Project-ID 352015383 – SFB 1330 A2, B1, and C4.